**SPAC: A Python Package for Spatial Single-Cell Analysis of Multiplex Imaging**

Fang Liu[1], Rui He[2], Andrei Bombin[3], Ahmad B. Abdallah[4], Omar Eldaghar[4], Tommy R. Sheeley[4], Sam E. Ying[4], George Zaki[1*]

**Affiliations**

[1]Frederick National Laboratory for Cancer Research, United States

[2]Essential Software Inc., United States

[3]Axle Informatics, United States

[4]Purdue University, United States

*Corresponding author

**Summary**

Multiplexed immunofluorescence microscopy captures detailed measurements of spatially resolved, multiple biomarkers simultaneously, revealing tissue composition and cellular interactions in situ among single cells. The growing scale and dimensional complexity of these datasets demand reproducible, comprehensive and user-friendly computational tools. To address this need, we developed SPAC (**SPA**tial single-**C**ell analysis), a Python-based package and a corresponding shiny application within an integrated, modular SPAC ecosystem (Liu et al., 2025) designed specifically for biologists without extensive coding expertise. Following image segmentation and extraction of spatially resolved single-cell data, SPAC streamlines downstream phenotyping and spatial analysis, facilitating characterization of cellular heterogeneity and spatial organization within tissues. Through scalable performance, specialized spatial statistics, highly customizable visualizations, and seamless workflows from dataset to insights, SPAC significantly lowers barriers to sophisticated spatial analyses.

**Statement of Need**

Advanced multiplex imaging technologies, such as CODEX (Goltsev et al., 2018), MxIF (Gerdes et al., 2013) , CyCIF (Lin et al., 2018), generate high dimensional dataset capable of profiling up to dozens of biomarkers simultaneously. Analyzing and interpreting these complex spatial protein data pose significant computational challenges, especially given that high-resolution whole-slide imaging data can reach hundreds of gigabytes in size and contain millions of cells across extensive tissue areas. Currently, many spatial biology tools (e.g., Seurat(Hao et al., 2021), GraphST (Long et al., 2023), and bento (Mah et al., 2024), primarily address spatial transcriptomics and cannot directly handle multiplexed protein imaging data. Other specialized software such as SPIA (Feng et al., 2023), Giotto (Dries et al., 2021), Squidpy (Palla et al., 2022), and SCIMAP (Nirmal et al., 2024) provides valuable capabilities tailored for spatial protein analyses. However, these tools lack sufficient flexibility and customization options necessary to meet the diverse scalable analysis and visualization needs of non-technical users.

To address this gap, we developed the SPAC Python package and the web-based SPAC Shiny application, which together enhance analytical capabilities through intuitive terminology, optimized computational performance, specialized spatial statistics, and extensive visualization configurations. Results computed using the SPAC Python package are stored as AnnData objects, which can be interactively explored in real time via the SPAC Shiny web application, enabling researchers to dynamically visualize data, toggle annotations, inspect cell populations, and compare experimental conditions without requiring extensive computational expertise.

Specifically, SPAC uses biologist-friendly terminology to simplify technical AnnData concepts. In SPAC, "cells" are rows in the data matrix, "features" denote protein expression levels, "tables" contain transformed data layers, "associated tables" store spatial coordinates or dimensional reductions (e.g., UMAP embeddings), and "annotations" indicate cell phenotypes, experimental labels, slide identifiers, and other categorical data.

To address real-time scalability challenges in analyzing large multiplex imaging datasets (exceeding 10 million cells), SPAC enhances computational efficiency by over 5x by integrating optimized numerical routines from NumPy's compiled C-based backend. Traditional visualization methods, such as seaborn, were computationally inefficient at this scale. SPAC's modified routines reduce visualization processing times from tens of seconds to a few seconds for generating histograms, box plots, and other visualizations involving millions of cells.

SPAC introduces specialized functions that enhance conventional spatial analyses. For example, SPAC implements a specialized variant of Ripley's L statistic to evaluate clustering or dispersion between predefined cell phenotype pairs, a "center" phenotype relative to a "neighbor" phenotype. Unlike generalized Ripley's implementations (e.g., Squidpy), SPAC explicitly distinguishes phenotype pairings and employs edge correction by excluding cells located near the region's borders within the analytical radius, mitigating edge-effect biases and enhancing statistical reliability. Furthermore, SPAC supports flexible phenotyping methods, accommodating both manual and unsupervised approaches tailored to diverse experimental designs and biological questions. It also implements efficient neighborhood profiling via a KDTree-based approach, quantifying the distribution of neighboring cell phenotypes within user-defined distance bins. The resulting three-dimensional array, capturing the local cellular microenvironment, is stored in the AnnData object and supports dimensionality reduction methods like spatial UMAP (Giraldo et al., 2021). This enhances comparative analysis and visualization of complex spatial relationships across multiple slides and phenotype combinations.

SPAC provides customizable visualization methods, leveraging Plotly's interactive capabilities for dynamic exploration of spatial data. Interactive spatial plots allow users to toggle of features (e.g., biomarkers) and multiple annotations simultaneously, while a pin-color option ensures consistent color mapping across analyses. These designs help researchers intuitively explore spatial relationships by switching between different cell populations and identify patterns before performing detailed quantitative analyses. In addition, SPAC supports comparative visualization, such as overlaying manual classifications with unsupervised clustering or comparing spatial distributions across experimental conditions or treatments. It also enhances core analytical functions (e.g., nearest neighbor computations using SCIMAP's spatial distance calculations) by integrating extensive visualization configurations, including subgroup analyses, subset plots, and faceted layouts, allowing tailored visual outputs for various experimental contexts and research questions.

**Structure and Implementation**

The SPAC package is available at https://github.com/FNLCR-DMAP/SCSAWorkflow and can be installed locally via conda. It includes five modules that streamline data processing, transformation, spatial analysis, visualization, and utility functions. The data utils module standardizes data into AnnData objects, manages annotations, rescales and normalizes features, and performs filtering, downsampling, and essential spatial computations (e.g., centroid calculation). The transformation tools module employs clustering algorithms (e.g., Phenograph, UTAG (Kim et al., 2022), KNN), dimensionality reduction, and normalization methods (batch normalization, z-score, arcsinh) to translate high-dimensional data into biological interpretation. The spatial analysis module offers specialized functions like spatial interaction matrices, Ripley's L statistic with edge correction, and efficient KDTree-based neighborhood profiling. It supports stratified analyses, capturing spatial signatures of cell phenotypes. The visualization module provides interactive and customizable visualizations, allowing dynamic exploration of spatial relationships and comparative visualization across experimental conditions. The utils module includes helper functions for input validation, naming conventions, regex searches, and user-friendly error handling to ensure data integrity.

All SPAC modules are interoperable, forming a cohesive workflow (Figure 1). By adopting the AnnData format, SPAC ensures broad compatibility with existing single-cell analysis tools, produces high-quality figures, and facilitates easy export for external analyses. SPAC adheres to enterprise-level software engineering standards, featuring extensive unit testing, rigorous edge-case evaluation, comprehensive logging, and clear, context-rich error handling. These practices ensure reliability, adaptability, and easy-of-use across various deployment environments, including interactive Jupyter notebooks, analytic platforms (e.g., Code Ocean (Code Ocean, 2019), Palantir Foundry (Palantir Technologies, 2003)), and real-time dashboards such as Shiny. The shiny interactive real-time dashboard for SPAC is available at [https://github.com/FNLCR-DMAP/SPAC_Shiny](https://github.com/FNLCR-DMAP/SPAC_Shiny). Emphasizing readability and maintainability, SPAC provides a versatile and enhanced analytical solution for spatial single-cell analyses. To date, SPAC has been used in the analysis of over 8 datasets with over 30 million cells across diverse studies (Keretsu et al., 2022).

**Acknowledgements**

We thank our collaborators at the National Cancer Institute Frederick National Laboratory for their invaluable feedback and contributions, and the single-cell and spatial imaging communities for providing the open-source packages and resources essential to our work.

**Figure 1. An overview of the SPAC Workflow.** The schematic presents an integrated pipeline for spatial single-cell analysis. Segmented cell data with spatial coordinates from various imaging platforms are ingested, normalized, clustered and phenotyped, and analyzed spatially to assess cell distribution and interactions while maintaining consistent data lineage.

**1**

**Single Cell Data**
- HALO
- Visiopharm
- QuPath
- IMC
- CSV files

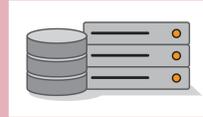
Single Cell Data

**2**

**Data Ingestion**
- Add new annotations
- Subsampling
- Manual phenotyping
- Pin color mapping

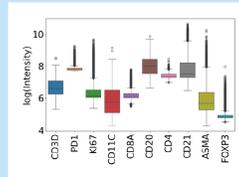
Manual Phenotyping via Phenotypes Codes

**3**

**Visualization**
- Histogram
- Boxplot
- Heatmap
- Hierarchical clustering
- Relational heatmap
- Sankey plot
- Scatter plot

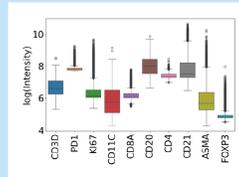
Distribution of Annotations or Features

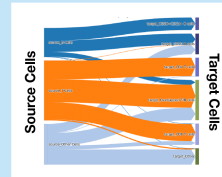
Relation Analysis Between Annotations

**4**

**Data Preprocessing**
- Arcsinh transformation
- Quantile normalization
- Rescaling
- Z-score normalization
- Batch correction

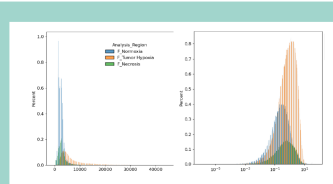
Feature Expression Before vs. After Z-score Normalization

**5**

**Clustering and Phenotyping**
- Phenograph clustering
- UTAG clustering
- tSNE/UMAP
- Spatial UMAP

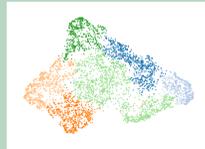
Clustering and Phenotyping Identification

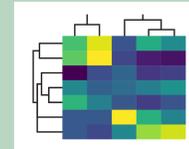
Hierarchical Clustering and Feature Expression

**6**

**Spatial Analysis**
- Interactive spatial plot
- Nearest neighbor
- Cluster interaction matrix
- Neighborhood enrichment
- Ripley's L
- Spatial hypothesis testing

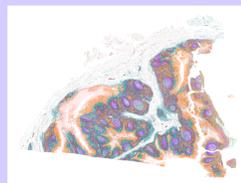
Spatial Distribution

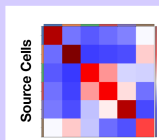
Neighborhood Enrichment

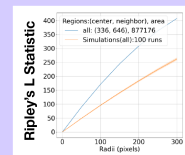
Ripley's L